 \documentclass[final,3p,times]{elsarticle}

 \usepackage{graphics}
 \usepackage{graphicx}

\usepackage{amssymb}

\usepackage{ulem}

\journal{Journal of Molecular Spectroscopy}

\begin{document}

\begin{frontmatter}



\title{The Cologne Database for Molecular Spectroscopy, CDMS, in the Virtual Atomic 
       and Molecular Data Centre, VAMDC}


\author[Koeln]{Christian P. Endres\fnref{PA}}
\fntext[PA]{Present address: Max-Planck-Institut f{\"u}r extraterrestrische Physik, 
            Giessenbachstrasse 1, 85748 Garching, Germany}
\author[Koeln]{Stephan Schlemmer}
\author[Koeln]{Peter Schilke}
\author[Koeln]{J{\"u}rgen Stutzki}
\author[Koeln]{Holger S.P. M{\"u}ller\corref{cor}}
\ead{hspm@ph1.uni-koeln.de}
\cortext[cor]{Corresponding author.}

\address[Koeln]{I.~Physikalisches Institut, Universit{\"a}t zu K{\"o}ln, 
   Z{\"u}lpicher Str. 77, 50937 K{\"o}ln, Germany}

\begin{abstract}

The Cologne Database for Molecular Spectroscopy, CDMS, was founded 1998 to provide in 
its catalog section line lists of mostly molecular species which are or may be observed 
in various astronomical sources by means of (usually) radio astronomical means. 
The line lists contain transition frequencies with qualified accuracies, intensities, 
quantum numbers, as well as further auxilary information. They have been generated from 
critically evaluated experimental line lists, mostly from laboratory experiments, 
employing established Hamiltonian models. Seperate entries exist for different isotopic 
species and usually also for different vibrational states. As of December 2015, the number 
of entries is 792. They are available online as ascii tables with additional files 
documenting information on the entries.

The Virtual Atomic and Molecular Data Centre, VAMDC, was founded more than 5 years ago 
as a common platform for atomic and molecular data.  This platform facilitates exchange 
not only between spectroscopic databases related to astrophysics or astrochemistry, 
but also with collisional and kinetic databases. A dedicated  infrastructure was developed 
to provide a common data format in the various databases enabling queries to a large variety 
of databases on atomic and molecular data at once.

For CDMS, the incorporation in VAMDC was combined with several modifications on the generation 
of CDMS catalog entries. Here we introduce related changes to the data structure and the data 
content in the CDMS. The new data scheme allows us to incorporate all previous data entries but 
in addition allows us also to include entries based on new theoretical descriptions. 
Moreover, the CDMS entries have been transferred into a mySQL database format. 
These developments within the VAMDC framework have in part been driven by the needs 
of the astronomical community to be able to deal efficiently with large data sets obtained 
with the Herschel Space Telescope or, more recently, with the Atacama Large Millimeter Array.

\end{abstract}


\end{frontmatter}




\section{Introduction}
\label{introduction}

The unambiguous assignment of a spectral line feature observed in space to a specific 
molecular or atomic species and its quantitative interpretation, e.g. in terms of column 
densities or relative abundances, depends decisively on the availability of rest frequencies 
and intensity information with sufficient accuracy. Such information is usually generated 
from laboratory data, sometimes supplemented by astronomical observations for rest 
frequencies and quite often by quantum chemical calculations for intensity information.
The reliability of the derived frequencies and intensities depends on the quality of the 
Hamiltonian model, the accuracies of the rest frequencies and of the intensity information, 
but also very much on the proper judgement of the line frequency accuracies. We will 
discuss important aspects on this topic in section~\ref{creating_entries}.

The CDMS\footnote{http://www.astro.uni-koeln.de/cdms/} \cite{CDMS_1,CDMS_2} was founded 
in the year 1998 to provide such information in its catalog section following 
the example of and complementing the JPL catalog\footnote{http://spec.jpl.nasa.gov/} 
\cite{JPL-catalog_1998} because of limited activities in the JPL catalog in recent years. 
The emphasis in the CDMS catalog is put on molecular species of interest for astrophysics 
or astrochemistry which have been or may be observed by radio astronomical means. 
The catalog also contains entries for fine structure transitions of atoms, such as C, 
$^{13}$C, C$^+$, $^{13}$C$^+$, N$^+$, O, Al etc. In selected cases, rovibrational 
transitions are available not only in the submillimeter (sub-mm) or far-infrared 
(FIR) region, such as for C$_3$ or C$_3$O$_2$, but sometimes also in the mid-IR, 
such as for CH$^+$, CCH, HCO$^+$, and N$_2$H$^+$. Some of the species are also important 
for studies of planetary atmospheres, including that of Earth, for example HCN, H$_2$CO, 
PH$_3$, CH$_3$CN, and SO$_2$. Individual entries are created for different isotopologues 
as far as they may be relevant for astronomers. In addition, transitions of different 
vibrational states are usually provided in different entries as well; they may be in 
one single entry for simple diatomics, such as NaCl and KCl, or strongly coupled 
vibrational states. The number of entries approaches 800 at the end of 2015; an 
up-to-date list is available online\footnote{http://www.astro.uni-koeln.de/cdms/entries}. 
These entries, together with those in the JPL catalog, cover a large fraction of the 
almost 200 molecules detected in the interstellar medium (ISM) or in circumstellar 
envelopes (CSEs)\footnote{See, e.g., http://www.astro.uni-koeln.de/cdms/molecules.} 
of late type stars with minor isotopologues or excited vibrational states. In fact, the CDMS 
and JPL catalogs were instrumental for the analyses of spectral recordings obtained, 
e.g., with the Herschel Space Telescope. In addition, data collections, such as the 
Leiden Atomic and Molecular DAtabase (LAMDA) \cite{LAMDA_2005}, CASSIS\cite{CASSIS_2011}, 
and splatalogue \cite{splat_2006,splat_2010}, draw heavily on the entries of the 
CDMS and JPL catalogs. Other databases, such as GEISA \cite{GEISA_2009,GEISA_2016} and 
HITRAN \cite{HITRAN_2012}, use many data from the CDMS and JPL databases, in particular 
in their pure rotational parts.

The line list entries are generated from spectroscopic parameters, their uncertainties, 
and their correlation coefficients. These values are obtained from experimental data 
employing established Hamiltonian models. The experimental data usually stem from 
laboratory spectroscopic investigations, but data from astronomical observations 
can also be used as far as appropriate. In most cases, the SPFIT/SPCAT program suite 
\cite{spfit_1991} is used for fitting and prediction of molecular spectra. Other programs 
have been used on occasion, details are provided in section~\ref{Hamiltonians}.

In the early years of the CDMS, considerable effort had been put on terahertz data of 
light hydrides, such as entries on H$_2$D$^+$, HD$_2^+$, CH, CH$^+$ and its isotopologues, 
CH$_2$, NH, ND, NH$_2$, $^{15}$NH$_3$, OH$^+$, H$_2$O$^+$, SH$^+$, and H$_2$S, H$_2$Cl$^+$, 
and ArH$^+$ along with their isotopologues. The aim was to support missions such as the 
Herschel Space Telescope \cite{Herschel_2010}, especially with respect to its 
high resolution Heterodyne Instrument for the Far-Infrared, HIFI, \cite{HIFI_2010} 
and the Stratospheric Observatory For Infrared Astronomy, and SOFIA, 
\cite{SOFIA_ES_2012} with its German REceiver for Astronomy at Terahertz frequencies, 
GREAT, \cite{GREAT_2012}. \textit{Herschel} in particular has expanded our knowledge 
in the field of terahertz astronomy tremendously \cite{Herschel-results_2012}, 
in particular with respect to hydrides. One of the main goals were low-energy 
rotational transitions of H$_2$O and its isotopologues \cite{interstellar_water_review_2013}. 
Large scale molecular line surveys were another important goal. 
They led to the detection of several hydrides in Galactic or extragalactic sources, 
the latest detection being ArH$^+$ \cite{ArH+_diff-ISM_2014,det_ArH+_2013}, a tracer 
of the almost completely atomic ISM, that was also detected in an extragalactic source 
recently with ALMA \cite{extragal_ArH+_2015}. Important discoveries with SOFIA/GREAT 
include the detection of SH \cite{SH_det_2012} and of the ground state rotational 
transition of \textit{para}-H$_2$D$^+$; the latter observation benefitted from 
earlier laboratory work \cite{H2D+_1-0_2008}, see also section~\ref{line_issues}. 
Terahertz observation are possible to some extent even from the ground, 
as demonstrated, e.g., by the early observations of OH$^+$ \cite{OH+_det_2010} 
and SH$^+$ \cite{SH+_det_2011} with the Atacama Pathfinder EXperiment, 
APEX, \cite{APEX_2006}.

The focus of new or updated entries in the CDMS has shifted somewhat with the advent 
of the Atacama Large Millimeter Array, ALMA. Due to the very high sensitivity and high 
spatial resolution, which reduces confusion caused by multiple source components 
in the telescope beam, interferometers are superior to single dish telescopes 
in detecting complex organic molecules as exemplified by the detection of 
\textit{iso}-propyl cyanide (\textit{i}-C$_3$H$_7$CN) as the first branched 
alkyl compound \cite{i-PrCN_det_2014} whose detection was only possible 
because of fairly recent laboratory work \cite{i-PrCN_rot_2011}. 
Transitions belonging to vibrationally excited states are easier to detect, 
e.g. highly excited HNC toward the carbon-rich late-type star CW~Leonis 
\cite{ALMA_vib-HNC_2013}; such transitions provide deeper insight into 
excitation mechanisms in general and insight into the dust formation zone 
in the particular case of late-type stars. Interferometric observations of 
late-type stars provide considerably more detail than single dish observations and 
can lead to the detection of new species, such as the recent detection of TiO$_2$ and 
the radio astronomical detection of TiO in the circumstellar envelope of the O-rich 
AGB star VY~Canis Majoris \cite{TiO_TiO2_det_2013}. TiO$_2$ was also observed with 
ALMA during science verification observations \cite{TiO2_ALMA_2015}. Interferometric 
observations will also be very beneficial for the investigations of extragalactic 
sources, such as the unnamed foreground galaxy in the direction of the quasar PKS~1830-211 
\cite{ALMA_PKS1830-211_2014}, in which also ArH$^+$ was detected \cite{extragal_ArH+_2015}.

Radio astronomical spectra are often analyzed under the assumption of local thermodynamic 
equilibrium (LTE). This is usually a good assumption in warm and dense parts of molecular 
clouds, but deviations from LTE may be considerable at lower temperatures, e.g. in dark 
clouds, or in less dense regions, especially in the diffuse ISM. In these cases it is 
often necessary to take into account collisional processes with H$_2$ and He in the denser 
ISM, and/or with H and electrons in the more diffuse ISM. A review on collisional processes 
in the ISM has been published fairly recently \cite{collisions_review_2013}.
Accurate data are needed for a plethora of molecules to avoid misinterpretations. 
An interesting and important aspect was the detection of seemingly more HNC than the more 
stable HCN. These results were obtained by assuming the somewhat similar molecular structure 
would lead to similar collisional rates with He and H$_2$. In the absence of collisional 
data for HNC, those of HCN were used instead. Initial calculations of HNC with He and later 
with H$_2$ \cite{HNC-H2_2011} revealed that the collisional rates are in fact very different 
and the column densities of HNC had to be revised to values lower than those of HCN in all 
instances.

Collisional rates are available from laboratory experiments for some systems. 
In many cases, however, such data are not available, and often it may be difficult 
to obtain such data. Resorting to quantum chemical calculations is a common 
alternative. But the calculation of collisional processes by quantum chemical means 
with high accuracy is demanding even nowadays. 
Therefore, calculations are often restricted to He as the collider and rates 
of collisions with H$_2$ are often estimated from these. Also, these calculations 
consider mostly collisons with small molecules, which play a more important role 
in cold or less dense environments than complex organic molecules. Neverteless, the 
need for collisional data involving more complex molecules has become inceasingly 
apparent in recent years. The complex absorption and emission spectrum of methyl 
formate (CH$_3$OCHO) in the GBT PRIMOS survey of Sagittarius~B2(N) between $\sim$0.3 
and $\sim$50~GHz is an interesting case because the observed low energy transitions 
sample the less dense envelope of this source, and their intensity distribution, 
including weak maser activity in some lines, could not be explained by an LTE model 
\cite{collisions_MeFo_2014}.

The need to access an increasing number of accurate molecular spectroscopy data 
together with an increasing number of accurate collisional data has become 
apparent in recent years. An early effort to link molecular spectroscopy data 
(here from the CDMS and JPL catalogs) with collisional data in a virtual 
observatory compliant environment was undertaken within the BASECOL collisional 
data base \cite{basecol_2006,basecol_2013}. 
Eventually, this effort led to the Virtual Atomic and Molecular Data Centre 
(VAMDC)\footnote{http://www.vamdc.eu/; http://www.vamdc.org/} \cite{VAMDC_2010}, 
an infrastructure which links further collisional databases, molecular spectroscopy 
databases in other frequency regions (e.g. databases such as HITRAN \cite{HITRAN_2012} 
or ExoMol \cite{ExoMol_2012}), atomic spectroscopy databases, and also kinetic 
databases such as KIDA \cite{KIDA_2012} and UMIST \cite{UMIST_2013}. The CDMS has 
participated in the VAMDC from the beginning, and one main part of this article, 
section~\ref{CDMS_VAMDC_content}, describes changes in the content of the CDMS 
catalog that were required for reliable data exchange between the various VAMDC 
partners, but also changes that have been made to serve the users of the CDMS catalog.

As a consequence of supporting VAMDC's infrastructure, the CDMS has undergone a 
technical transformation process. The data are stored in a relational database 
(mysql) and standards, such as query languages to access the database, and 
output formats, which are developed within VAMDC will be supported. This 
process comes along with an improvement of the provided data in terms of their 
physical meaning and description. Besides simplifying the use and access of the 
data, one of the goals is also to make the data origin more transparent and 
enable others to reproduce the data. This aspect is described in 
section~\ref{CDMS_VAMDC_infrastructure}.


\section{Considerations for creating CDMS catalog entries}
\label{creating_entries}

The generation of CDMS catalog entries requires several steps. The first one is the decision 
that the spectrum of a (mostly) molecular species is or may be of interest for the community 
that uses the CDMS catalog or the decision that an existing entry requires update because of 
the need and availability of additional, usually new, data of sufficient impact. 
Collecting available experimental line lists may be anything from very straightforward 
to time-consuming, as can be the case for the subsequent steps. These are the generation 
of Hamiltonian models, evaluation of intensity-related aspects, such as dipole moment 
components and partition functions, and finally the generation of a documentation 
which provides background information on the entry.


\subsection{Aspects related to experimental line lists}
\label{line_issues}

Line lists of rotational data may consist of just a single line, but may also be extensive, 
especially in cases of molecules with a complex spectrum, for example because of fine 
or hyperfine structure splitting, such as the NH$_2$ radical. A large data set from 
diverse sources was assembled in Ref.~\cite{NH2_rot_1999}. In cases like this one, 
it is particularly important to evaluate the correctness of reported line frequency 
uncertainties and estimate values in cases in which no uncertainties have been reported. 
Since this process is not independent of the choice of Hamiltonian parameters, 
an iterative process may be required. In the end, it is not only desirable that 
the entire experimental line list has been reproduced within ascribed uncertainties 
on average, but this should also be tested for reasonably well defined subsets. 
Details are again available for NH$_2$ \cite{NH2_rot_1999}. Another example of a 
CDMS related publication is the fit of radio-frequency to terahertz data of several 
NO isotopologues plus heterodyne IR data for the main species \cite{NO_rot_2015}.

The inclusion of rovibrational or rovibronic data in a fit is particularly important 
if the data known with microwave accuracy are very limited. Only the $J'' = 0$ (i.e., 
$J = 1 - 0$) rotational transition frequencies of CH$^+$, $^{13}$CH$^+$, and CD$^+$ 
have been published thus far \cite{CH+_rot_2010}. These data were combined subsequently 
with extensive rovibronic data of the $A - X$ electronic transitions of four 
isotopologues to evaluate rotational and rovibrational transition frequencies 
\cite{CH+_parameter_2010}. The prediction of rotational transitions of the main 
isotopologue were accurate enough for analysis of emission spectra taken with 
the moderate resolution PACS instrument on board the \textit{Herschel} satellite 
toward a disk around Herbig~Be star HD~100546 which extended to $J'' = 5$ 
\cite{CH+_up_to_6-5_2011}. Another example is argonium. Extensive rotational and 
rovibrational data are available for $^{40}$ArH$^+$ and $^{40}$ArD$^+$ because 
$^{40}$Ar (from the radioactive decay of $^{40}$K) is by far the most abundant argon 
isotope on Earth. In the Solar atmosphere and in the interstellar medium, $^{36}$Ar 
is most abundant, with $^{38}$Ar lower by a factor of around five, and $^{40}$Ar being 
largely negligible. The only published data with microwave accuracy involving isotopologues 
containing $^{36}$Ar or $^{38}$Ar are the $J'' = 0$ transition frequencies of $^{36}$ArD$^+$ 
and $^{38}$ArD$^+$ \cite{ArD+_isos_rot_1983}. Combining these data with the extensive 
rotational and rovibrational data for $^{40}$ArH$^+$ and $^{40}$ArD$^+$ yielded CDMS 
entries that facilitated the identification of the $J'' = 0$ and 1 transitions of 
$^{36}$ArH$^+$ in emission in the Crab Nebula \cite{det_ArH+_2013}, the $J'' = 0$ 
transition of $^{36}$ArH$^+$ in absorption in the diffuse medium toward several 
continuum sources, of the same transition of $^{38}$ArH$^+$ toward Sagittarius~B2(M) 
\cite{ArH+_diff-ISM_2014}, and of the $J = 1 - 0$ transitions of $^{36}$ArH$^+$ 
and $^{38}$ArH$^+$ in a $z = 0.89$ foreground galaxy in absorption toward the 
quasar PKS~1830-211 \cite{extragal_ArH+_2015}. Several cationic molecules, such as 
CH$^+$ or argonium, occur mostly or exclusively in the diffuse interstellar medium 
because in the dense medium they are readily destroyed by reaction with H$_2$. 
Excitation of such molecular cations by collision with H$_2$ can not be invoked to 
explain observation of some of the cations in emission under specific circumstances; 
instead, excitation through collisions with electrons, as evaluated, e.g., in 
Ref.~\cite{e-exc_CH+_ArH+_2016}, or with H atoms can explain such observations. 
In the case of sparse rotational data, accurate ground-state combination differences 
from IR spectra can be very useful also, such as recently determined for isotopic species 
of H$_3^+$ \cite{H3+_isos_IR_2016}.

Experimental lines from astronomical observations can also be very useful, in particular 
for molecules that are difficult to synthesize in the laboratory. In some cases, initial 
assignments were based on astronomical data only, such as in the cases of C$_5$N$^-$ 
\cite{C5N-_det_2008} and C$_3$H$^+$ \cite{C3H+_det_2012}. Laboratory rotational data 
have been obtained for the latter cation in the meantime \cite{C3H+_rot_2014,C3H+_rot_2015}, 
in great part because of the dispute concerning the carrier of the lines. Data from 
astronomical observations can even be important if they only supplement laboratory data 
and if they have lower accuracies, such as in the case of extensive \textit{Herschel}-HIFI 
data on the cyclic SiC$_2$ molecule \cite{SiC2_param_2012}.

The correctness, i.e., the accuracy and precision, of a reported transition frequency 
is an important aspect for molecules with small line list. In the case of rotational 
ground state transitions of H$_2$D$^+$ \cite{H2D+_1-0_2008} or CH$^+$ \cite{CH+_rot_2010} 
errors of $\sim$60~MHz were revealed with respect to earlier reports. 
Frequency errors may also be found in the course of astronomical observation, 
such as the case of the lowest fine structure component lines of SH$^+$ near 
346~GHz \cite{param_SH+_2014}. Subsequent laboratory measurements provided improved 
accuracies for these lines as well as remeasurements of the fine structure component 
lines near 526~GHz along with two of the four hyperfine lines of the highest fine 
structure component of the $N = 1 - 0$ transition \cite{SH+_rot_2015}. Such errors are 
easier to identify in larger data sets, where, in addition, the omission of a single 
line has frequently a small effect.

The uncertainties of experimental lines in a fit are of major concern. Earlier publications 
frequently refrained from quoting any specific value, but this has changed considerably 
in recent years. Nevertheless, a recent paper quotes uncertainties of better than 50~kHz 
up to 500~kHz for a very large data set. This information, however, is not particularly 
useful if no uncertainties appear in the line list. Sometimes uncertainties of previously 
reported data get modified without justification and without apparent reason. 
We have found quite frequently that the reported transition frequency uncertainties 
were judged too conservatively or too optimistically. The uncertainties of the resulting 
spectroscopic parameters may be somewhat reasonable if so-called standard errors are reported. 
However, the uncertainties of the lines are still incorrect, and the lines may carry 
a too large or too small weight in more extensive line lists. It is always much better 
to try to evaluate appropriate uncertainties; this is often not too difficult if the line 
list is sufficiently large and the Hamiltonian model is complete.


\subsection{Hamiltonian models}
\label{Hamiltonians}

CDMS catalog entries are mostly generated employing Pickett's SPFIT/SPCAT program suite 
\cite{spfit_1991}, a versatile program developed to fit asymmetric rotors with multiple 
vibrational states or nuclear or electronic spins. The program evolved with time 
\cite{editorial_Herb-Ed,intro_JPL-catalog}. Extended treatment of symmetry and 
spin-statistics \cite{Dn_2004} and the introduction of Euler functions for fitting spectra 
of moderately rigid molecules, such as H$_2$O \cite{Euler_2005} were major extensions 
of the program. Special considerations exist for linear molecules and symmetric tops, 
especially for $l$-doubled states. Quantum numbers as well as the number of vibrational 
states can exceed 100 by far.

The choice of spectroscopic parameters to be used in a fit is often far from unique. 
This is particularly the case if the primary objective to include a reasonable parameter 
in a fit is its determination with significance. Such approach leads too easily to 
too large parameter sets in cases of complex spectra such as the rotation-tunneling 
spectrum of the lowest energy conformer of ethanediol \cite{eglyc_rot_2003} or the 
Coriolis interaction between $\varv_4 = 1$ and $\varv_6 = 1$ of ClClO$_2$ 
\cite{ClClO2_Cori_2002}. Searching at all stages of the fit among the reasonable 
spectroscopic parameters for the one that reduces the root mean squares (rms) error 
of the fit most appears to be a suitable strategy to keep the number of spectroscopic 
parameters in the fit small. The importance of the choice of reductions and 
representations have been discussed often, Ref.~\cite{aziridine_rot_2010} is an example 
in which other aspects (e.g. condition numbers) have also been discussed.

In multi-state fits, there are different ways to reduce the number of spectroscopic 
parameters similar to the familiar Dunham expansion for diatomic molecules. 
Such approaches can be applied to fitting of rotation-tunneling spectra, as in the cases 
of ethanediol \cite{eglyc_rot_2003}, H$_2$DO$^+$ \cite{H2DO+_param_2010}, and 
ethanethiol \cite{ROH_RSH_2016}. In the case of \textit{gauche}-C$_2$H$_5$SH, the number 
of spectroscopic parameters was reduced to 34 from 41 in a traditional two-state fit 
\cite{EtSH_rot_2014}.

The combination of diverse data pertaining to one molecular species can improve the 
spectroscopic parameters considerably. This is often not only beneficial from the 
molecular physics point of view, but also for the predictions to be used for radio 
astronomical observations; some examples are CH$^+$ \cite{CH+_parameter_2010}, 
H$_3$O$^+$ \cite{H3O+_rot_fitting_2009}, H$_2$DO$^+$ \cite{H2DO+_param_2010}, 
and PH$_3$ \cite{PH3_param_2013}.

A careful choice of spectroscopic parameters has led in some cases to predictions that 
turned out in later investigations to be considerably better than those based on original 
publications, as in the cases of oxirane \cite{oxirane_FIR_2012} and CHD$_2$CN 
\cite{D-MeCN_rot_2013}. Technical aspects of a fitting program may also lead to incorrect 
Hamiltonian parameters, as has happened for NaCN \cite{NaCN_param_2012}.

We point out that there are entries in the CDMS catalog which have not been generated with 
SPFIT/SPCAT. This applies, in particular, to molecules which display splitting caused by 
large amplitude motions such as the internal rotation of one or more CH$_3$ groups. Several 
programs are available, as described in a fairly recent review \cite{internal_rotor_review_2010}. 
Existing entries in the CDMS catalog include CH$_3$OH and $^{13}$CH$_3$OH \cite{MeOH_rot_1997}, 
dimethyl ether \cite{DME_rot_2009}, CH$_3$SH \cite{MeSH_rot_2012}, CH$_3^{18}$OH 
\cite{Meo-18-H_rot_2007}, and CH$_3$O$^{13}$CHO \cite{13C1-MeFo_rot_2001}.


\subsection{Dipole moments and intensities}
\label{dipole_moments}

Dipole moment components of a given molecule are important pieces of information to calculate 
the intensity of a rotational transition at a certain temperature, its line strength $S\,\mu^2$, 
where $S$ is the intrinsic line strength, and $\mu$ is the effective dipole moment component, 
or its Einstein $A$ value\footnote{http://www.astro.uni-koeln.de/cdms/catalog\#equations} 
(e.g. \cite{JPL-catalog_1998}). They can be determined from Stark effect measurements which 
have been quite common in times of Stark-modulation rotational spectroscopy, but are less common 
nowadays. Nevertheless, new measurements include TiO$_2$ \cite{TiO2_dip_2009}, 1,2-propanediol 
\cite{1-2-PD_FTMW_2009}, and 1,3-propanediol \cite{1-3-PD_FTMW_2009}. Occasionally, 
dipole moment components get reinvestigated, as in the cases of the lowest energy conformer of 
\textit{n}-propanol \cite{n-PrOH_rot_2010}, \textit{iso}-propyl cyanide \cite{i-PrCN_rot_2011}, 
and ethyl and vinyl cyanide \cite{VyCN_EtCN_dip_2011}. Such measurements usually improve 
older measurements considerable, but in some cases, one or more dipole moment components 
get revised considerably such as in the cases of \textit{iso}-propyl cyanide \cite{i-PrCN_rot_2011} 
and vinyl cyanide \cite{VyCN_EtCN_dip_2011}.

Stark effect measurements seek in most cases to resolve the individual Stark components as well 
as possible. An opposite approach does work, however. In the weak field limit, one seeks to keep 
all Stark components within a single and symmetric line and measures only a small Stark shift 
of some tens of kilohertz; this can lead to dipole moment components accurate at the percent 
level \cite{ClClO2_rot_1999}.

Quantum chemical calculations of dipole moment components are an important alternative for 
experimental determinations. In some cases, even selections of several molecules are studied 
to support the astronomical community \cite{dips_div_2009,dips_Si_u_P-comp_2013}. 
Such calculations \cite{transition-dips_rev_2014} are even more important for transition dipole 
moments of not so stable molecules because reliable intensity measurements may be difficult 
or even impossible.

Rotational, also know as centrifugal distortion, and other corrections may have to be taken into 
account in particular for rotational spectra of light hydrides or for molecules with large amplitude 
motions. They may also matter for rigid molecules in highly rotationally excited transitions. 
 In cases of mixing of states mediated by internal rotation, tunneling, or vibration-rotation 
interaction, sign choices of spectroscopic parameters, which describe the mixing, 
relative to those of the dipole moment components may have significant impact on the intensities 
of some transitions. Examples have been given for the lowest energy conformer of ethanediol 
\cite{eglyc_rot_2003} or H$_2$DO$^+$ \cite{H2DO+_param_2010}. Recently, intensity problems 
in the rotation-tunneling spectrum of ethanol in the 3~mm wavelength region became apparent 
by astronomical observations with ALMA and were, at least largely, resolved by a change of 
the sign of one dipole moment component \cite{ROH_RSH_2016}.


\subsection{Partition functions}
\label{partition_functions}

The partition function $Q$ is very important to evaluate intensities of molecular or 
atomic lines assuming thermal population or to derive column densities. In the usual case, 
in which entries have been generated with the SPCAT progam \cite{spfit_1991}, partition 
function values at selected temperatures are calculated directly by summation over 
the energy levels. These values, sometime supplemented by additional values at other, 
in particular lower or higher temperatures, are provided in the documentation file, 
see section~\ref{documentation_files}; a separate 
page\footnote{http://www.astro.uni-koeln.de/site/vorhersagen/catalog/partition\_function.html} 
lists the decadic logarithm of these values. The partition function can be separated to first 
order into a rotational part ($Q_{\rm rot}$) and a vibrational part ($Q_{\rm vib}$). 
Traditionally, the CDMS as well as the JPL catalog were usually restricted to ($Q_{\rm rot}$) 
only, i.e., they have provided partition function values for the ground vibrational state only. 
More complete partition function values have become available in more recent entries.

The Boltzmann constant used for evaluation of the partition function is about $2 \times 10^{-6}$ 
larger than the 2010 CODATA value \cite{CODATA2010}; this has a negligible effect on the 
calculated intensities. In addition, uncertainties in the partition function because of 
limitations in the accuracy of spectroscopic parameters, in particular the vibrational energies, 
see below, are usually much larger also, except for some diatomics.

If the ground state partition function values are determined by numerical summation over 
the energies of a particular species, it is important that the maximum $J$ (or $F$) and 
$K$ values have been chosen high enough to ensure convergence of the partition function 
values at 300~K. The SPCAT program calculates these values down to units of 10$^{-4}$. 
The CDMS entries have usually been generated applying a partition function value 
that is converged to that level or at least nearly so. In the case of the light CH radical, 
this is achieved at $F = 15$. In case of the still light HCN molecule, $J = 49$ is required. 
HC$_3$N and HC$_5$N require $J = 152$ and 298, respectively. However, this criterion is rather 
strict because the CDMS catalog provides the decadic logarithm of the intensity at 300~K 
given to four digits after the decimal point, thus a $Q$(300) value too low by a factor 
of 1.000115 affects the decadic logarithm of the intensity at best in the last digit given. 
In the case of HC$_5$N, this means that $J = 206$ is already sufficient for proper 
calculation of the decadic logarithm of the intensity to all quoted digits in most cases. 
Easing the accuracy requirements to 1\,\% and 10\,\% reduces the maximum $J$ value to be 
considered to 147 and 103, respectively.

Calculating the partition function values up to 300~K or even higher by summation over the 
energies may not be always feasible, in particular for very large molecules or for molecules 
with a very complex Hamiltonian. Treating the molecule as a rigid rotor is a very efficient 
alternative; a better compromise could be to calculate the partition function by summation 
over the energies up to a certain temperature and correct the rigid rotor approximation for 
higher temperatures by extrapolation from lower temperature deviations. These approaches 
are good or at least reasonable as long as the molecule is not too floppy.

Care is advised if the partition function is calculated by summation over the energies 
if there are relatively large distortion parameters in the Hamiltonian. For example, if 
$D_K$ or $L_K$ is the highest order purely $K$-dependent distortion term in an asymmetric 
rotor close to the prolate limit, higher $K$ levels will be lower in energy than lower ones 
eventually, and the energies may eventually even turn negative. This is one important reason 
why maximum $J$ and $K$ quantum numbers should not be chosen much higher than needed.

Assuming thermal population, excited vibrational states may even be populated at temperatures 
just barely larger than those of dark clouds. At 300~K, a vibration with vibrational energy 
of 480~cm$^{-1}$ or 691~K is down in population by a factor of 0.1 with respect to 
the ground state. At twice the energy or half the temperature, the population is down 
by a factor of 0.01. Larger organic molecules may have low-lying vibrations (torsional modes 
or bending modes) which may well be below 100~cm$^{-1}$ or 144~K. 
In addition, they may also have more than one low-lying conformation. 
In the case of, e.g., ethanol and $n$-propanol, the combined vibrational and 
conformation factors at 150~K were evaluated to be 2.96 and $\sim$5.2 if the respective 
data set is restricted to the ground vibrational state of the lowest energy conformer only 
\cite{ROH_RSH_2016}. Vibrational corrections to the partition function, and hence to 
the column density, are difficult to evaluate because in particular low-lying vibrational 
states (in the FIR region) are often not known or only known with limited accuracy because 
the transitions are weak, only rather low-resolution measurements were made, or the 
measurements were not made in the gas phase. Additional complications may arise from 
incorrectly reported experimental data, such as in the case of vinyl cyanide, in which the 
energy of $\nu _{10}$ was revised from 788 to 561~cm$^{-1}$ \cite{VyCN_IR_2015}. 
Quantum-chemical calculations on vibrational properties at diverse levels of theory are 
quite common nowadays. High-level calculations are often very demanding on larger molecular 
systems, and the resulting accuracy may still be limited. A combination of experimental and 
quantum-chemical data may be a good compromise in many cases. Experimentally, the anharmonic 
fundamental vibrations are determined to a varying extent. Quantum-chemical calculations 
provide in most instances the harmonic vibrations. Detailed information on the anharmonicity 
is rarely available. Nevertheless, estimating contributions from overtone and combination states 
via the harmonic oscillator approximation provide better estimates of the vibrational contributions 
than the restriction to known vibrational levels. Various hybrid ways of evaluating vibrational 
contributions can be made. In Ref.~\cite{MeCN_v8_le_2_2015}, the rotational part of the partition 
function was converged up to room temperature. Vibrational contributions up to $\varv _8 = 3$ 
at almost 1100~cm$^{-1}$ were taken into account directly; the truncation error was estimated 
to be 1.408\%  at 296~K via the harmonic oscillator approximation. A different type of 
mixed approach was taken for three isotopologues of methyl formate \cite{Q_MeFo_2014}. 
Torsional contributions up to $\varv _{\rm t} = 6$ were evaluated from data up to 
$\varv _{\rm t} = 2$, and contributions from the other vibrations were estimated by 
the harmonic oscillator approximation. For larger molecules, a fully classical calculation 
of the partition function may be the most reasonable way. This was done, e.g., for the acetone 
entry in the JPL catalog, a fairly heavy molecule with two very low lying torsional modes 
and a rather low CCC bending mode.

For the calculation of astrophysical line intensities, one should keep in mind that 
a significant uncertainty arises from the fact that in real molecular clouds high 
vibrational states are not necessarily in LTE with the gas temperature, and that, 
therefore, the partition function calculated under this assumption will be erroneous. 
In many cases, vibrational states are excited by absorption of IR radiation, 
and their excitation temperature is therefore determined by the effective radiation 
temperature of the IR, which is not necessarily equal to the kinetic temperature 
which will determine the LTE of the vibrational ground state through collisions. 
The details will depend on the exact source structure, and would need to be determined 
by exact radiative transfer calculations.


\subsection{Documentation files}
\label{documentation_files}

For each catalog entry, there is a documentation which provides basic information on 
the content of the entry. These are the chemical formula, the name (possibly alternative 
names), the symmetry of the (usually ground) electronic state, and possibly vibrational 
information, the tag to identfy the species, the number of lines (i.e. transition frequencies), 
the highest quantum number, and the highest frequency. The references, from which 
the experimental transition frequencies were taken, are listed usually completely, 
though some early entries may list references only in part. Information on the treatment 
of the data may also be given, such as omission of data, judgment of uncertainties, 
or information on the Hamiltonian model. Other information includes estimates on the 
reliability of the entry with respect to the needs of astronomers, partition function 
values, often also information on limitations of the partition function (e.g., the 
usual limitation to the ground vibrational state, or the consideration of vibrational 
states), rotational constants, dipole moment components, among other information. 
Links to special entries, for example with or without HFS, special spin states, etc., 
may also be given. The consideration of non-trivial spin-statistics is generally 
mentioned as well.


\section{Content of the VAMDC-CDMS}
\label{CDMS_VAMDC_content}

The main focus of the CDMS database is providing accurate transition frequencies 
for astronomical use. This biases the selection of molecular and atomic species and 
the selection of vibrational states. Usually, experimental data reported in the 
literature have been fit to an effective Hamiltonian. Subsequently, the derived 
parameters are used to calculate transition frequencies, intensities as well as 
state energies.

The data are organized in entries which are identified by a unique, sequential 
identifier which replaces the formerly used species tag.  Different atomic and 
molecular species are assigned to different entries, but one entry comprises at 
most data stemming from the same compilation. This means the data from one 
entry have been fit together and assures thereby that the data in themselves 
are consistent. Data from the same compilation have been distributed into two 
or more entries in several cases. This may be the case if more than one 
vibrational state have been involved.

Each entry contains information on radiative transitions, state energies, partition 
functions, origin of the data and a documentation. Data are only provided if transition 
frequencies are known accurately enough to be useful for astronomical observations; 
this means if the rest frequencies are known better than one MHz and on average better 
than 100~kHz. Older versions of entries have been archived in either CDMS version since 
the start of the VAMDC project. This allows one to review data retrieved earlier and 
assures backward traceability.

\subsection{Documentation of entries}

Documentation information continues to be available for each entry. The main difference 
with respect to the single documentation page in the old CDMS implementation is 
that certain pieces of information are available in separate navigation items. 
These include a brief description of the data used in the fit and estimates on 
the reliability of the provided data sets. 
Basic molecular constants, such as dipole moments and rotational constants, as well as 
some parameters which have been used for the compilation of the data are given. 
The files which have been used to prepare the entry are usually provided in order 
to allow traceability and thereby to improve the quality assurance. 
Links to cited resources are provided as well as links to other data repositories, 
such as ChemSpider\footnote{http://www.chemspider.com}, NIST Chemistry 
WebBook\footnote{http://webbook.nist.gov}, and databases connected to the VAMDC network.

\subsection{Radiative Transitions}

The listings of radiative transitions represents the central part of the 
database. Rest frequencies, uncertainties, intensities at 300~K, Einstein $A$ 
values and quantum numbers are given for each listed transition. Upper and 
lower states are linked in order to provide information on the energy (in 
cm$^{-1}$) as well as on the degeneracy of the states involved in the 
transition.  Quantum numbers are now reported in physical meaningful and 
commonly used quantities by applying the standard introduced by VAMDC (see 
section below). This improves not only the readablility of the quantum numbers,
but is also of importance to match state information among different databases.
The old ascii format, which is based on the output of the Pickett's SPCAT program, 
will still be available and stored internally in order to simplify quality assurence
and bug tracking, as well as to support third party programs which may rely
on the traditional format.

\subsection{State energy listing}

The databases now provide rovibrational state energy listings for each entry. In most 
cases, the band origin is not as accurately known as the transition frequencies. 
Absolute values are accurate only if rovibrational transitions have been included 
in the fit. Usually, the energy of the band center is taken from the literature 
and absolute uncertainties of one or several wavenumbers are common. 
The relative values within a vibrational level are, in contrast, as accurate as 
specified, because their values are obtained from the fitted experimental 
transition frequencies and are thus the basis on which transition frequencies 
are calculated. Although their absolute values may be known only to about a 
wavenumber, the state information might still be a reliable basis for 
collisional data. The state energy listing also constitutes the basis on which 
the partition functions are calculated.

\subsection{Present considerations of evaluation of partition functions}

Partition functions are provided for 110 temperatures between $1~$K and $1000~$K. 
They are calculated as a sum over all states of a specific species listed in the database. 
If the data of one species are distributed over more than one entry, which is regularly 
the case if data for many vibrational states are reported, all states of that specie 
are linked to each of these entries in order to provide the most complete partition 
function possible based on the available data (covering all reported states). 

\begin{equation}
 Z(T) = \sum_v \sum_{rot-states} g_I e^{-E / k_B T}  \mbox{\qquad , {\it v} in CDMS}
\end{equation}

\noindent
with $g_I$ the spin-statistical weight, and $E$ the energy of the respective 
rovibrational level.

Since the state information reported in the database may not be sufficiently 
comprehensive to allow the calculation of reliable partition functions even 
at low temperatures for a number of molecules, in particular if low-lying 
torsional states are present, a routine to calculate the partition function 
based on the harmonic approximation has been introduced in the database tools 
and is used to provide more accurate values in such cases if the result 
of the summation is known to be insufficient. At present, only the harmonic 
oscillator approximation for asymmetric top molecules is included: 

\begin{eqnarray}
 Z_{rot}(T) &=& g_{tot} T^{1.5} / \sigma  \sqrt{( \pi (k / h)^{3} / (A  B  C ) } \\
 Z_{vib}(T) &=& \prod_v (1 - e^{- E(v) / kT})^{-1} \\
 Z(T) &=& Z_{rot}(T)  Z_{vib}(T)
\end{eqnarray}

\noindent
where $A$, $B$, and $C$ are the rotational constants, $\sigma$ is the symmetry number, 
and $g_{tot}$ is the total spin statistical weight.

\subsection{References}

Large efforts have been made to improve citations to original data used in the 
database and to supplement the documentation. VAMDC rules prescribe that 
references have to be given to all provided values. Thus, each experimental 
frequency reported in the database includes the reference pointing to the 
original publication. Nevertheless, some entries remain for which explicit 
references to experimental frequencies are still missing and are only included 
in the documentation.

\section{Infrastructure of the VAMDC-CDMS}
\label{CDMS_VAMDC_infrastructure}

A new infrastructure for atomic and molecular data has been developed within the VAMDC 
framework. CDMS has been migrated over the last couple years to VAMDC and is now available. 
Standard web-techniques have been applied allowing easy integration of database services 
into any computer program or webservice. Data access and retrieval is based on the 
http protocol. Queries to the databases can be formulated already in any web-browser 
within an URL-string and a large variety of methods is available for almost any computer 
language to implement this protocol and thus gain access to VAMDC databases. 
VAMDCs standard data output and data exchange format is VAMDC-XSAMS, a modified version 
of the International Atomic Energy Agency's XML Schema for Atoms, Molecules and Solids 
(XSAMS) \cite{VAMDC_2010}.

\subsection{Access Protocols and Query/Retrieval Language}

VAMDC uses the web-service protocol VAMDC-TAP for data access services. That means, 
queries use GET and POST requests to HTTP/HTTPS URLs and the syntax of the URL is 
defined by the protokol. VAMDC-TAP is a variant of the Table Access Protocol (TAP) 
defined by IVOA (International Virtual Observatory Alliance) \cite{VAMDC_2010}. 
Each VAMDC-TAP service (database) can return results of queries in VAMDC-XSAMS, 
supports synchronous data-queries and provides a description of its availability 
and its service capabilities as specified in TAP by reference to the Virtual 
Observatory Support Interfaces (VOSI) standard. The query language, which is 
supported by all VAMDC databases, is VAMDC SQL sub-set \#2 (VSS2). It defines 
how queries requesting specific data from VAMDC databases have to be formulated. 
It is similar to the well-known database language SQL. An example VAMDC-TAP request 
to CDMS querying all data for the CO molecule could be carried out the following way: 

\begin{verbatim}
http://cdms.ph1.uni-koeln.de/cdms/tap/sync?
 REQUEST=doQuery
 &LANG=VSS2
 &FORMAT=XSAMS
 &QUERY=SELECT+ALL+WHERE+MoleculeStoichiometricFormula='CO'
\end{verbatim}

The same request can be sent to any other VAMDC database, just by exchanging the base 
URL appropriately.

\subsection{Data Models and XML Schema}

The standard model VAMDC-XSAMS is structured in sections specifying species, physical processes, 
sources, methods, functions, and environments. References to data sources and information on 
methods used to generate the data is linked to provided data values and are strict requirements 
to allow traceablility. Physical states of atoms, molecules, ions are described as unambigously 
and meaningful as possible. Information, such as statistical weights and a reference to the state 
specifying the energy origin can be attached, as in the following example, which specifies the 
$J=0$ ground state of the CO molecule:

\begin{verbatim}
<MolecularState stateID="SCDMS-9086558">
  <MolecularStateCharacterisation>
    <StateEnergy methodRef="MCDMS-6489" 
      energyOrigin="SCDMS-9086558-origin-83">
      <Value units="1/cm">0.0</Value>
    </StateEnergy>
    <TotalStatisticalWeight>1</TotalStatisticalWeight>
    <NuclearStatisticalWeight>1</NuclearStatisticalWeight>
  </MolecularStateCharacterisation>

  <Case xsi:type="dcs:Case" caseID="dcs" 
    xsi:schemaLocation="http://vamdc.org/xml/xsams/1.0/cases/dcs ../../cases/dcs.xsd">
    <dcs:QNs>
      <dcs:ElecStateLabel>X</dcs:ElecStateLabel>
      <dcs:v>0</dcs:v>
      <dcs:J>0</dcs:J>
    </dcs:QNs>
  </Case>
</MolecularState>
\end{verbatim}

Physical processes are defined in terms of states using a reference to initial and final state 
as shown in the following example, which describes the rotational transition $J=1-0$ of CO 
in the vibrational ground state: 

\begin{verbatim}
<RadiativeTransition id="PCDMS-R15140649" 
  process="excitation">
  <EnergyWavelength>
    <Frequency methodRef="MCDMS-6428">
      <SourceRef>BCDMS-1681</SourceRef>
      <Value units="MHz">115271.2018</Value>
      <Accuracy>0.0005</Accuracy>
    </Frequency>
    <Frequency methodRef="MCDMS-6488">
      <Value units="MHz">115271.2021</Value>
      <Accuracy>0.0001</Accuracy></Frequency>
  </EnergyWavelength>
  <UpperStateRef>SCDMS-9086559</UpperStateRef>
  <LowerStateRef>SCDMS-9086558</LowerStateRef>
  <SpeciesRef>XCDMS-83</SpeciesRef>
  <Probability>
    <TransitionProbabilityA>
      <Value units="1/cm">7.20360334988e-08</Value>
    </TransitionProbabilityA>
    <Multipole>E2</Multipole>
  </Probability>
  <ProcessClass>
    <Code>rota</Code>
  </ProcessClass>
</RadiativeTransition>
\end{verbatim}

\begin{figure}
    \includegraphics[width=15cm]{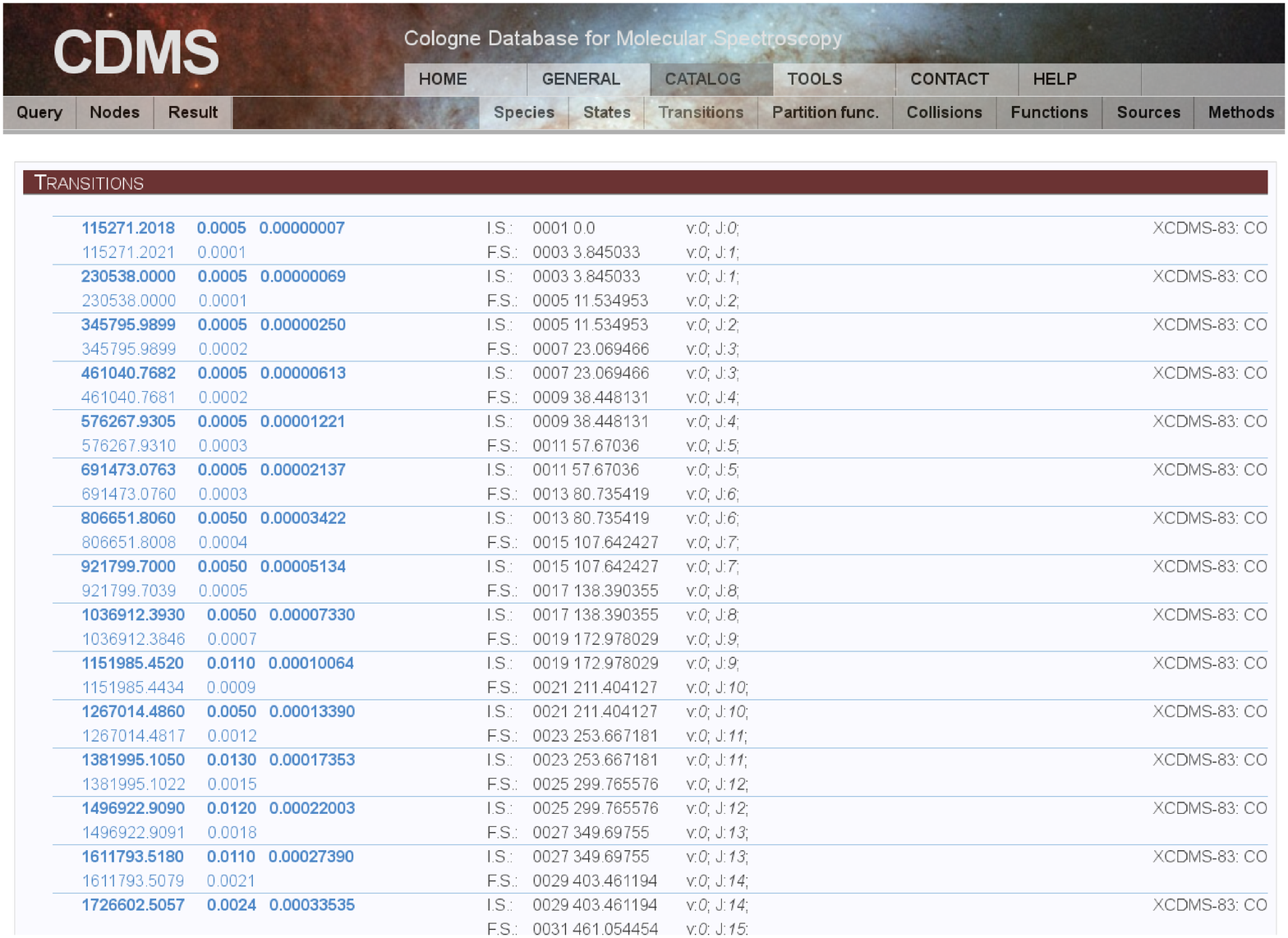}
    \caption{CDMS-Portal - Query page. The page displays information on transitions of 
        the $^{12}$C$^{16}$O molecule as obtained from CDMS. The data itself is obtained 
        as VAMDC-XSAMS file and transformed by the portal into the form shown above. 
        Frequencies [in MHz], uncertainties [in MHz], Einstein~$A$ values [in s$^{-1}$] are 
        provided in the first three columns. Experimental and calculated values are shown 
        in separate rows. The recommended values are shown in bold. References will be shown 
        if the mouse cursor is moved over the displayed value. The next columns contain 
        state information (statistical weight, energy [in cm$^{-1}$] and quantum numbers). 
        Lower state information is displayed in the upper row, whereas upper state 
        information is provided in the lower row.
        }
    \label{fig_portal_trans}
\end{figure}

As VAMDC-XSAMS is an XML schema, transformation into any other data format and 
extraction of data of a user's interest is easy and will also be supported by VAMDC 
libraries. An example is shown in Fig.~\ref{fig_portal_trans}. The result of a 
query requesting transitions of the CO molecule via VAMDC-TAP obtained in VAMDC-XSAMS 
is transformed by the CDMS portal into HTML code.

\subsection{Definition of quantum numbers}

An important aspect of the data exchange format VAMDC-XSAMS is the definition of 
quantum numbers. A simple, but precise definition of quantum numbers allowing unambigous 
description of states is key to allow automated cross-correlation of data between 
various databases. Therefore, different sets of quantum numbers have been defined 
reflecting the commonly used labeling schemes for all different groups of molecules, 
such as closed shell diatomic molecules (dcs), open shell asymmetric top molecules (asymos), 
open shell linear molecules (lpos), for example. Each set contains a sequence of quantum 
numbers following closely the commonly used notations. Vibrational quantum numbers 
are accompagnied by the an identifyer of the normal mode associated with it, various 
parity and symmetry labels are distinguished, and total (or intermediate) angular momentum 
quantum numbers include an identifier of the nuclear spin associated with it 
(e.g. $F$, nuclearSpinRef = "N"). The following example gives the definition for the 
$J_{K_a,K_c} = 1_{1,0}$, $F$ = 0, $s$ inversion state of NH$_2$D:

\begin{verbatim}
<Case xsi:type="asymcs:Case" caseID="asymcs" 
   xsi:schemaLocation="http://vamdc.org/xml/xsams/1.0/cases/asymcs ../../cases/asymcs.xsd">
  <asymcs:QNs>
    <asymcs:ElecStateLabel>X</asymcs:ElecStateLabel>
    <asymcs:vibInv>s</asymcs:vibInv>
    <asymcs:J>1</asymcs:J>
    <asymcs:Ka>1</asymcs:Ka>
    <asymcs:Kc>0</asymcs:Kc>
    <asymcs:F nuclearSpinRef="N">0</asymcs:F>
  </asymcs:QNs>
</Case>
\end{verbatim}

\subsection{Web-services and libraries}

Currently, there are several web-services, programs and libraries which give 
users access to the CDMS database. A common web-portal for data retrival has been 
developed by VAMDC\footnote{www.vamdc.eu/portal}.  It allows users to perform queries 
to all data and databases in the network and includes basic displaying 
capabilities.  In addition, a new portal for CDMS has been created 

\begin{figure}
    \includegraphics[width=15cm]{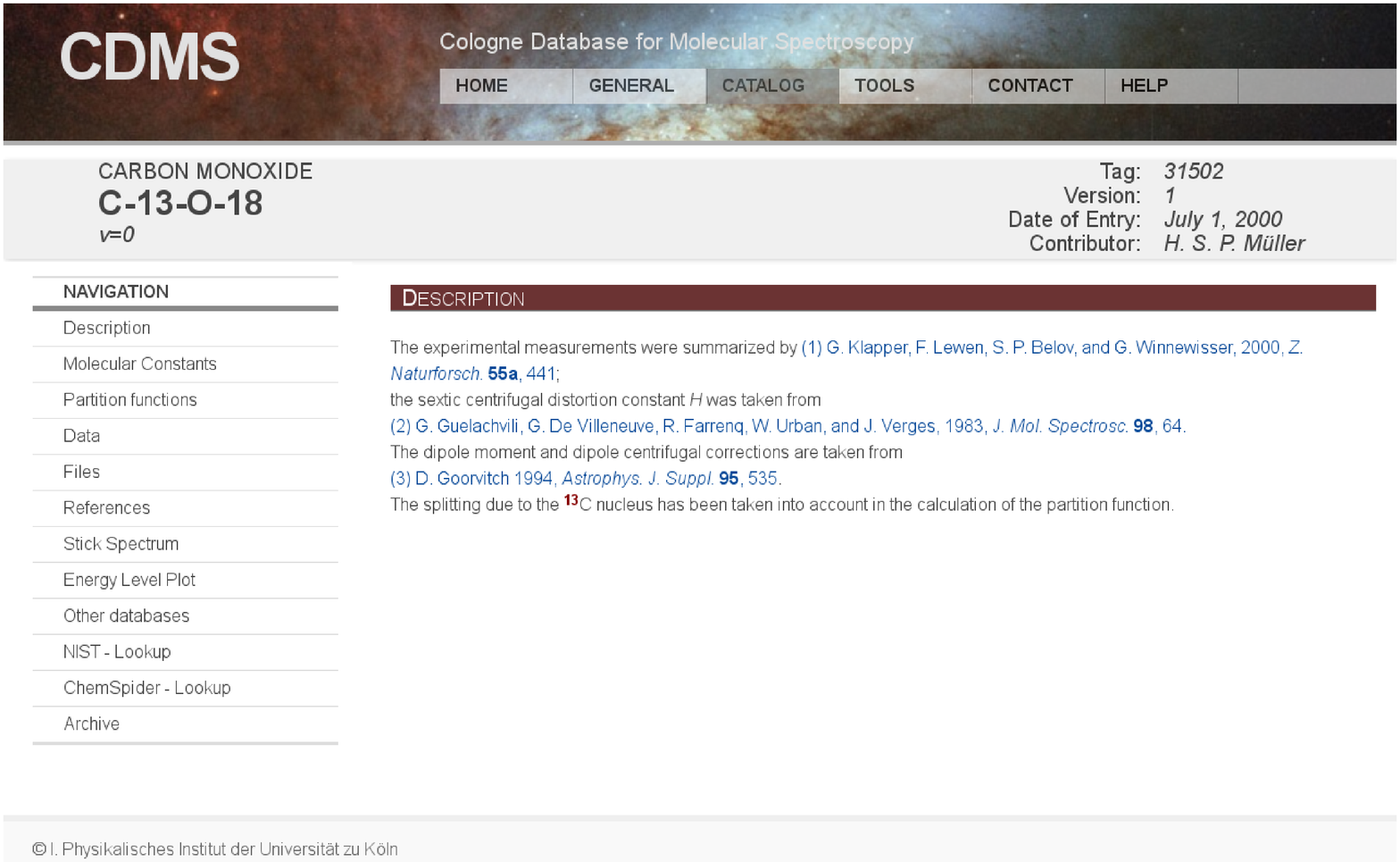}
    \caption{CDMS-Portal - Documentation page of the $^{13}$C$^{18}$O entry. The page provides
        basic information on the entry like the date of the entry, version, contributor and
        a brief description. Further information, such as molecular constants, references, 
        files, and links to other databases can be accessed through the navigation bar on the left.}
    \label{fig_portal_doc}
\end{figure}

\textbf{http://cdms.ph1.uni-koeln.de/cdms/portal} , 

\noindent
focused on the needs of the user community of the CDMS and JPL databases. 
Additional information, which cannot be exchanged within the XSAMS document, 
is implemented here, such as a detailed documentation (see Fig.\ref{fig_portal_doc}), 
files used in the fit and specific output formats (catalog file format of SPCAT).

Compared to the old CDMS portal, the new version provides extended filter capabilities. 
Querying all data data for a specific entry (e.g. CO, $\varv = 0$), for a specific 
isotopologue (e.g. $^{13}$CO) or for all isotopologues of a specific molecule (e.g. CO) 
is possible. Detailed information on states, transitions (see Fig.\ref{fig_portal_trans}) 
and partition functions are displayed and further restrictions can be posed to limit 
the number of displayed states to states within a specific energy range or transitions 
to a specific frequency range.  Transitions and states with and without 
resolved hyperfine structure as well as with and without information on 
the nuclear spin isomer is shown and filters can be applied accordingly.

A number of software tools are provided by 
VAMDC\footnote{http://www.vamdc.org/activities/research/software/} which can be applied 
to query and process data. SpectCol, for example, allows one to combine spectroscopic 
data with collisional data (e.g. from Basecol). Data can be converted into other output 
formats such as radex and csv. Several conversion (XSAMS Converter) and visualisation 
(SpecView) tools can be downloaded after registration. Development of external 
software packages is also supported. A python-library (vamdclib) has been developed 
for this purpose and has been successfully implemented into the XCLASS software package 
for CASA. It allows one to query and process data via the VAMDC database network and 
to store the data into a local sqlite database. 
XCLASS \cite{xclass_2015} uses the CDMS/JPL database to fit 
astrophysical line spectra in the LTE approximation.  It is available for the CASA 
platform\footnote{http://www.astro.uni-koeln.de/projects/schilke/myXCLASSInterface}, 
and its database provides the partition function in 100 steps between 1 and 1000~K, 
in order to correctly determine the column densities of very low excitation absorption lines.


\section{Summary and outlook}
\label{Outlook}

Over the last years, the CDMS content has been enlarged substantially, and the database 
has been subject to fundamental infrastructural improvements which makes it now part 
of the VAMDC database framework. Both measures were necessary steps to make CDMS fit 
for the use in the ALMA era. The focus of creating further new or updating existing 
entries will be concerned with species of interest for observations with ALMA or 
other interferometers. These include larger organic molecules and isotopic or 
vibrational satellite data of smaller organic molecules along with various radicals, 
ions, or stable molecules which may be detected in the circumstellar envelopes of 
young or late-type stars or in external galaxies. We welcome external contributions, 
in particular for complex and extensive data sets or for entries generated with 
programs differenr from SPFIT/SPCAT.

The migration of CDMS into VAMDC has been almost completed. The old version of CDMS 
will be available for some time to serve users who depend on the old ascii format. 
Based on the large amount of data coming from today's observations automated queries 
of CDMS alone or in combination with other databases using the http protocol are now 
possible and encouraged. All users will benefit from the various additional features 
the database entries as well as the new web portal have. CDMS will further been improved 
with the aim to keep its role as a very important resource for the interpretation of 
observations in the millimeter and sub-mm regimes.



\section*{Acknowledgements}

We thank the German Ministry of Science (BMBF) for support through contract 05A11PK3. 
We acknowledge in addition funding by the Deutsche Forschungsgemeinschaft (DFG) through 
the collaborative research grant SFB~956 ''Conditions and Impact of Star-formation'', 
project areas A6, B3, B4, and C3 and through grant Schl~341-15/1.








\end{document}